\documentclass[a4paper,11pt]{article}
\usepackage{pos}
\usepackage{multicol}
\usepackage{wrapfig}
\usepackage{caption}
\usepackage{subcaption}
\title{Three-loop master integrals for H+jet production at N$^3$LO: Towards the non-planar topologies}

\author[a,b]{Dhimiter D. Canko}
\author*[c]{Nikolaos Syrrakos}

\affiliation[a]{Institute of Nuclear and Particle Physics, NCSR "Demokritos",\\ Agia Paraskevi 15310, Greece}
\affiliation[b]{Department of Physics, University of Athens,\\Zographou 15784, Greece}

\affiliation[c]{Technical University of Munich, TUM School of Natural Sciences, Physics Department, James-Franck-Straße 1, 85748 Garching, Germany}
            
\emailAdd{jimcanko@phys.uoa.gr}
\emailAdd{nikolaos.syrrakos@tum.de}

\abstract{We discuss the recent progress that has been made towards the computation of three-loop non-planar master integrals relevant to next-to-next-to-next-to-leading-order (N$^3$LO) corrections to processes such as H+jet production at the LHC. We describe the analytic structure of these integrals, as well as several technical issues regarding their analytic computation using canonical differential equations. Finally, we comment on the remaining steps towards the computation of all relevant three-loop topologies and their application to amplitude calculations.}

\FullConference{16th International Symposium on Radiative Corrections: Applications of Quantum Field Theory to Phenomenology (
RADCOR2023)\\
28th May - 2nd June, 2023\\
Crieff, Scotland, UK\\
Preprint Number: TUM-HEP-1463/23\\}

\renewcommand{\hookAfterAbstract}{%
\par\bigskip
\textsc{ArXiv ePrint}:
\href{https://arxiv.org/abs/2307.08432}{2307.08432}
}

\begin{document}
\maketitle

\section{Introduction}
With the commencement of the LHC Run 3, increased data samples and higher collision energy will allow us to probe the nature of fundamental interactions to unprecedented precision \cite{Dainese:2019rgk}. It falls upon the theory community of particle physics to provide predictions of collider observables at such a level of precision, that will ensure the optimal exploitation of LHC data. To be more specific, considering the wealth of currently available data, as well as the expected high luminosity upgrade, N$^3$LO corrections will become essential in order to perform phenomenological studies at the 1\% level at the LHC \cite{Caola:2022ayt}. 

A vital ingredient of such precise theoretical predictions are the multi-loop scattering amplitudes. Their computation lies at the heart of the calculation of a hard scattering cross section for the production of the final state particles we are interested in. 

From a more technical standpoint, when considering the computation of a multi-loop amplitude, one needs to confront two fundamental challenges:
\begin{enumerate}
    \item Expressing the amplitude in terms of a (relatively) small set of so-called master integrals.
    \item Evaluating these master integrals. 
\end{enumerate}


The task we wish to undertake is the computation of N$^3$LO corrections for processes such as H+jet production in the heavy top quark-mass limit \cite{Wilczek:1977zn, Shifman:1978zn, Inami:1982xt}. The same master integrals involved in this process can also be used to study Z+jet production at N$^3$LO. Working in $d=4-2\epsilon$ dimensions to regulate UV and IR divergences, the loop amplitudes are computed as a Laurent expansion in $\epsilon$. An N$^3$LO calculation requires one-loop amplitudes up to $\mathcal{O}(\epsilon^4)$, two-loop amplitudes up to $\mathcal{O}(\epsilon^2)$ and three-loop amplitudes up to $\mathcal{O}(\epsilon^0)$. For H+jet production, the first step towards this goal was recently achieved in \cite{Gehrmann:2023etk}, with the computation of the one-loop and two-loop amplitudes to their required order in $\epsilon$. More recently, the relevant one-loop and two-loop amplitudes for Z+jet production were also computed to higher orders in $\epsilon$ \cite{Gehrmann:2023zpz}. At the three-loop level, results for all relevant planar master integrals were presented in \cite{DiVita:2014pza,Canko:2020gqp,Canko:2021xmn} and more recently some non-planar master integrals where computed in \cite{Henn:2023vbd}.

In this contribution we will address the latter of the previously mentioned challenges of a multi-loop amplitude calculation. In particular, we will focus on the master integrals that are relevant to the three-loop amplitudes for gg$\to$Xg and q$\Bar{\text{q}}$$\to$Xg, with X=H,Z. The remaining of this contribution is structured as follows. In section 2 we introduce the integral families relevant to the aforementioned scattering amplitudes and set up our notation. In section 3 we give a complete description of all relevant planar three-loop master integrals for an amplitude calculation. In section 4 we present analytic results for two non-planar topologies and discuss the ongoing effort to complete the full set of three-loop non-planar master integrals. We give our concluding remarks in section 5. 

\begin{table}
\begin{center}
\begin{tabular}{ |c||c c c|} 
 \hline
  & IR top sectors & R top sectors & MI \\ 
 \hline\hline
 PL & 3 & 17 & 291 \\ 
 NPL1 & 5 & 1 & 414 \\ 
 NPL1c34 & 3 & 1 & 328 \\ 
 NPL2 & 5 & 8 & 781\\ 
 NPL2c24 & 2 & 1 & 412\\
 \textbf{Total} & \textbf{18} & \textbf{28} & \textbf{2226}\\
 \hline
\end{tabular}
\end{center}
\caption{Integral families. R denotes the reducible and IR the irreducible top sectors.}
\label{tab: intfam}
\end{table}

\section{Integral families and kinematics}
To make our discussion more concrete and to set our notation for the following sections, let us properly define the kinematics of the process we are interested in. Let us take for example
\begin{equation}
     H(p_4) \rightarrow g_1(p_1) + g_2(p_2) + g_3(p_3).
\end{equation}
The external momenta satisfy $\sum_{i=1}^4 p_i=0,~p_4^2=q^2,~p_i^2=0$ for $i=1,2,3$, with $s_{ij} = (p_i+p_j)^2$. For convenience we will work with the following dimensionless variables
\begin{equation}
    x = \frac{s_{12}}{q^2}, \quad \quad  y = \frac{s_{13}}{q^2}, \quad \quad z = \frac{s_{23}}{q^2}. 
    \label{eqn:xyz}
\end{equation}

One way of computing scattering amplitudes is the so-called projector method \cite{Peraro:2019cjj, Peraro:2020sfm}, where one uses Lorentz invariance, gauge invariance and other symmetries of the problem at hand to decompose the amplitude in terms of an independent basis of tensor structures $\mathcal{T}_i$
\begin{equation}\label{eq:ampl}
    \mathcal{A}(p_1,p_2,p_3) = \sum_{i=1}^{N} \mathcal{F}_i(p_1,p_2,p_3)\mathcal{T}_i
\end{equation}
where $\mathcal{F}_i$ are the scalar form factors. One then proceeds by defining projector operators which admit the same tensorial decomposition as the amplitude
\begin{equation}\label{eq:proj}
    \mathcal{P}_j = \sum_{i=1}^{N} c_i^{(j)}(d;p_1,p_2,p_3)\mathcal{T}_i^{\dag}.
\end{equation}
Applying these projectors to the amplitude and summing over the polarizations of the external states singles out the specific form factor
\begin{equation}\label{eq:ff}
    \sum_{\text{pol}} \mathcal{P}_j \mathcal{A}(p_1,p_2,p_3) = \mathcal{F}_j(p_1,p_2,p_3).
\end{equation}
The form factors obtained through \eqref{eq:ff} are expressed in terms of a large number of scalar integrals. For X+jet production at three loops, these scalar integrals can be defined through
\begin{equation}\label{eq:intdef}
    \mathrm{I}_{a_{1}, ..., a_{15}} = \int \bigg(\prod_{l=1}^3 (-q^2)^{-\epsilon} e^{\gamma_E \epsilon} \frac{\mathrm{d}^d k_l}{i\pi^{d/2}}\bigg) \prod_{i=1}^{15} D_i^{-a_i}.
\end{equation}
We can organise these integrals into five so-called integral families, one planar and four non-planar. In table \ref{tab: intfam} we present an analysis of these five integral families, showing the total numbers of the top sectors, i.e. integrals that correspond to diagrams involving ten internal lines and four external legs, and the number of master integrals for each family. With R we denote the reducible top sectors that are not linearly independent but can contribute master integrals from their subsectors, and with IR we denote the irreducible top sectors that will yield top sector master integrals.

This analysis has been performed using \texttt{Kira} \cite{Maierhofer:2017gsa, Klappert:2020nbg}, a program that implements and automates Laporta's algorithm for integration-by-part identities for dimensionaly regulated Feynman integrals \cite{Chetyrkin:1981qh,Laporta:2000dsw}. Since the latest version of \texttt{KIRA} does not support an analysis on the available crossings of an integral family which would be needed for a full amplitude calculation, we show only numbers for the uncrossed families. 
\begin{figure} [t]
\centering
\includegraphics[width=6cm]{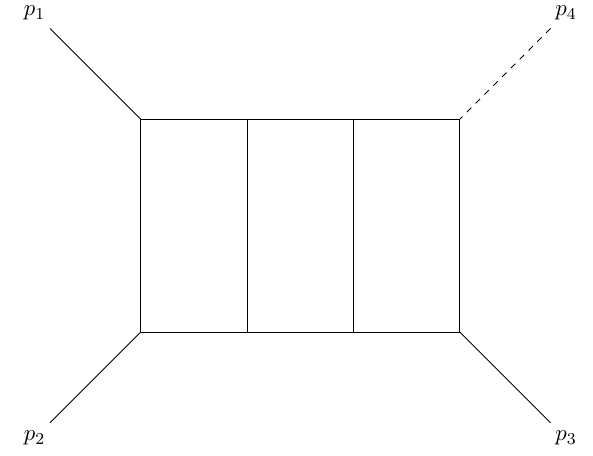}\\
\includegraphics[width=6.5cm]{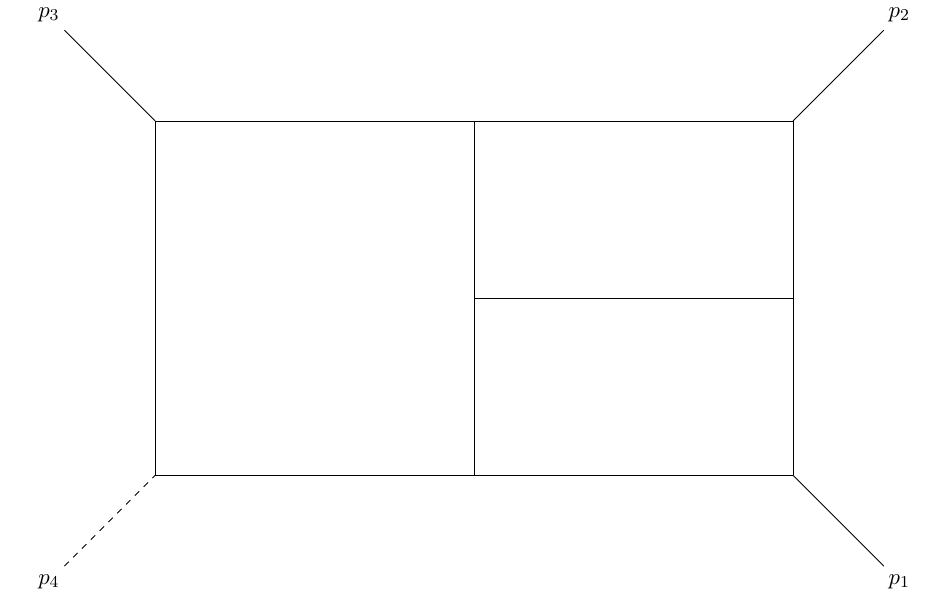}
\includegraphics[width=6.5cm]{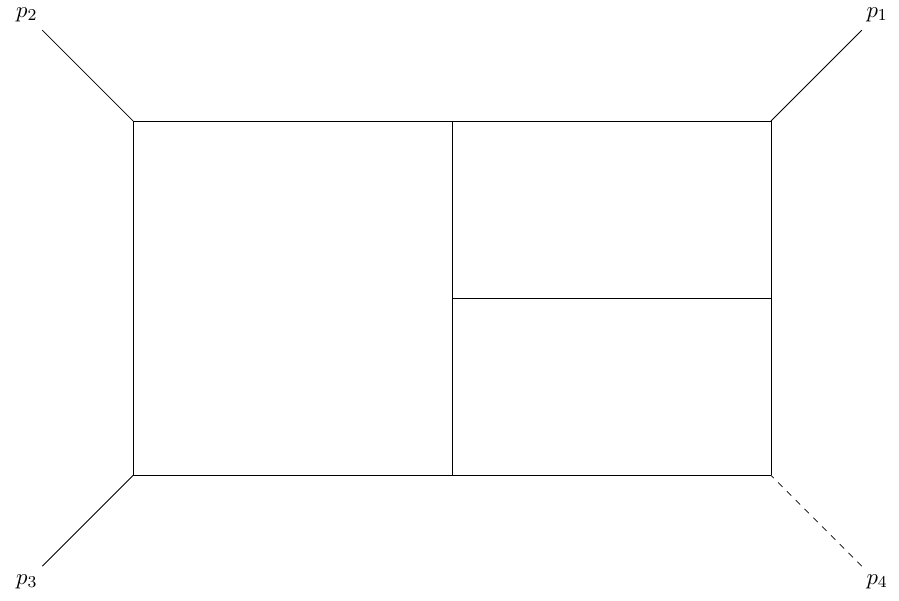}
\caption{Irreducible planar top sectors.}
\label{fig:planar_topgraphs}
\end{figure}

Our goal will be to consider each individual family of master integrals or individual top sectors and their subsectors, and construct canonical differential equations \cite{Kotikov:1990kg, Kotikov:1991hm, Kotikov:1991pm, Gehrmann:1999as, Henn:2013pwa} for them,
 \begin{equation}\label{eq:dlog_de}
    \mathrm{d}\vec{\mathrm{g}}=\epsilon  A \vec{\mathrm{g}} = \epsilon \sum_i B_i\, \mathrm{d}\log(\alpha_i)\vec{\mathrm{g}} ,
    \end{equation}
where we denote with $\vec{\mathrm{g}}$ the canonical basis of master integrals. The matrices $B_i$ consist of rational numbers and the arguments $\alpha_i$ of the $\mathrm{d}\log$ forms are known as letters. In general they can be rational or algebraic functions of the kinematic invariants. We will start from the Euclidean region defined through the kinematic invariants as
\begin{align}
        0<z<1,\quad 0<&y<1-z,\quad x=1-y-z,\\
        &\text{or}\nonumber\\
        0<z<1,\quad &0<x<1-z,
\end{align}
where all master integrals are real, and solve the differential equations analytically in terms of multiple polylogarithms \cite{Goncharov:1998kja},
\begin{align}
        G(a_1,a_2,\ldots, a_n;x) &= \int_0^x \, \frac{\mathrm{dt}}{t-a_1}G(a_2,\ldots, a_n;t)\\
        G(0,\ldots, 0;x) &= \frac{1}{n!}\log^n(x).
    \end{align}

\section{Planar topologies}
\begin{figure} [t]
\centering
\includegraphics[width=8cm]{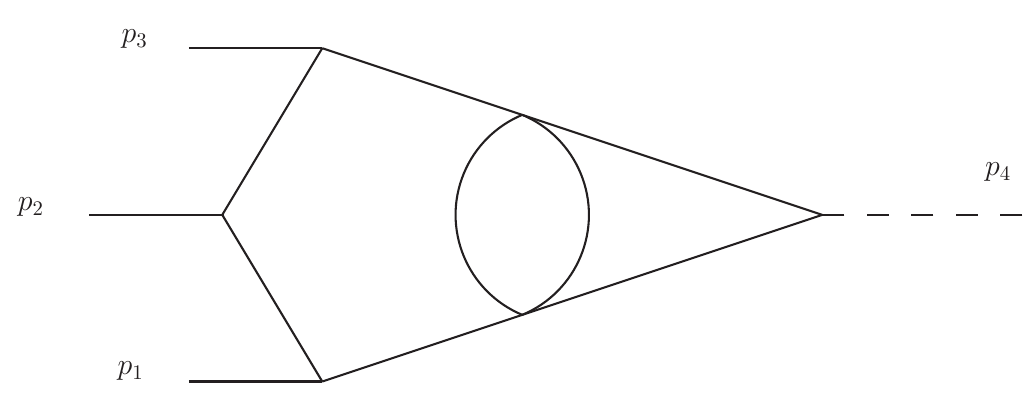}
\caption{Genuinely new three-loop master integral coming from one of the reducible top sectors.}
\label{fig:newmi}
\end{figure}

The 3 irreducible planar top sectors, shown in figure \ref{fig:planar_topgraphs}, along with their subsectors, were computed analytically in \cite{DiVita:2014pza,Canko:2020gqp,Canko:2021xmn}. For an amplitude calculation though, we also need to consider the contribution of the 17 reducible top sectors (see table \ref{tab: intfam}). We find that subsectors of these reducible scalar top sectors contribute 56 additional master integrals, bringing the total number of planar master integrals up to 291. Most of these new master integrals can be obtained through transformations of the kinematic invariants of the already known integrals. There is however a genuinely new master integral that appears, shown in figure \ref{fig:newmi}. 

In order to have results for all planar master integrals readily available for a three-loop amplitude calculation, we assembled a canonical basis of all 291 integrals and proceeded by deriving and solving canonical differential equations for them in the $y,~z$ variables. We verified that all planar master integrals satisfy the same six-letter alphabet as in the two-loop case,
\begin{equation}\label{eq:pl_letters}
    \{y,z,1-y,1-z,1-y-z,y+z\},
\end{equation}
and used familiar regularity constraints
\begin{equation}\label{eq:reg_b}
    \{y\to 0,\,y\to 1,\, y\to-z,\, z\to1\},
\end{equation}
to fix all boundary terms. We checked our results against all previously published ones and performed numerical checks against \texttt{pySecDec} \cite{Borowka:2017idc} for the additional master integrals. Furthermore, with the use of \texttt{HyperInt} \cite{Panzer:2014caa} and \texttt{PolyLogTools} \cite{Duhr:2019tlz}, we have obtained analytic expressions for all master integrals in all crossings that are needed for a full amplitude calculation.

\section{Non-planar topologies}

\begin{figure}[t] 
\centering
\begin{subfigure}[b]{0.5\textwidth}
    \includegraphics[width=\textwidth]{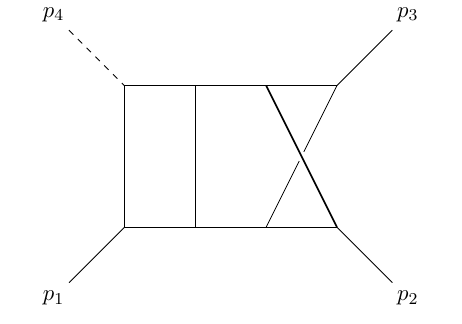}
    \caption{NPL2c24\_15055}
\end{subfigure}\hfill
\begin{subfigure}[b]{0.5\textwidth}
    \includegraphics[width=\textwidth]{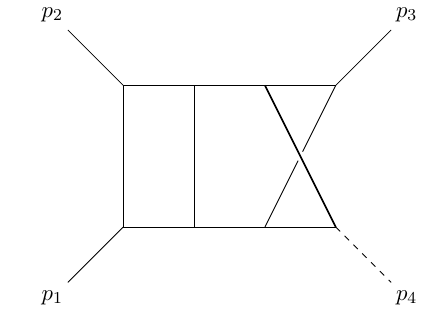}
    \caption{NPL2\_15055}
\end{subfigure}
\caption{The two ladder-like topologies considered here. The naming scheme used is \textit{family\_sectorID}.}
\label{fig:nonpltop}
\end{figure}

Moving on to the non-planar topologies, from table \ref{tab: intfam} we see that in total we need to consider 15 irreducible and 11 reducible top sectors, along with their subsectors. Experience has shown that, in order to construct canonical bases, it is more convenient to work with individual top sectors. As a starting point, we will consider the two irreducible ladder-like top sectors represented by the diagrams in figure \ref{fig:nonpltop}. Canonical bases and analytic results in terms of MPLs for these top sectors have been recently published in \cite{Henn:2023vbd}. Here we report on an independent calculation of these topologies. We will also comment on preliminary results for two non-planar top sectors that have not yet appeared in the literature. 

For topology (\textbf{a}) and its subsectors we found a total of 114 master integrals, with 4 of them at the top sector, using \texttt{Kira}. Respectively for topology (\textbf{b}) we find 150 and 4 master integrals. The next step involved the construction of canonical bases. To do so, we followed the same approach as in the planar topologies, i.e. using the \texttt{Mathematica} package \texttt{DlogBasis} \cite{Henn:2020lye} for sectors with up to 9 propagators and then doing a loop-by-loop analysis in $d=4$ dimensions on the maximal cut, to identify candidates with constant leading singularity. 

Verifying that the bases obtained with the methods as described above are indeed canonical, involves the derivation of their differential equations. This step relies heavily on IBP reduction. We have found that a standard use of \texttt{Kira}, i.e. giving a list of integrals that appear on the right-hand-side of the canonical differential equations and extracting their IBP reduction, is computationally very inefficient. To circumvent this issue, we have developed a framework based on \texttt{Reduze} \cite{vonManteuffel:2012np}, \texttt{Kira} and \texttt{Mathematica}, that allows us to directly reduce the canonical differential equations in terms of the canonical master integrals. The main steps are as follows:
\begin{enumerate}
     \item Generate the IBP identities with \texttt{Kira} for each top sector.
     \item Collect the integrals from the canonical basis and compute their derivatives with \texttt{Reduze}.
     \item Construct the  unreduced canonical differential equation with \texttt{Mathematica}. 
     \item Feed the system of IBPs and canonical differential equations to \texttt{Kira} and solve it as a user-defined system over finite fields.
\end{enumerate}
In terms of performance, it takes roughly 3.6h to generate $\sim \mathcal{O}(9\cdot10^7)$ IBPs per top sector. The most computationally intensive part is the last one, which required 11h and 19h to reduce three differential equations, one for each of the variables $(s_{12},\,s_{23},\,q^2)$, for each top sector respectively, on a machine using 50 CPU cores.

\begin{figure}[t] 
\centering
\begin{subfigure}[b]{0.5\textwidth}
    \includegraphics[width=\textwidth]{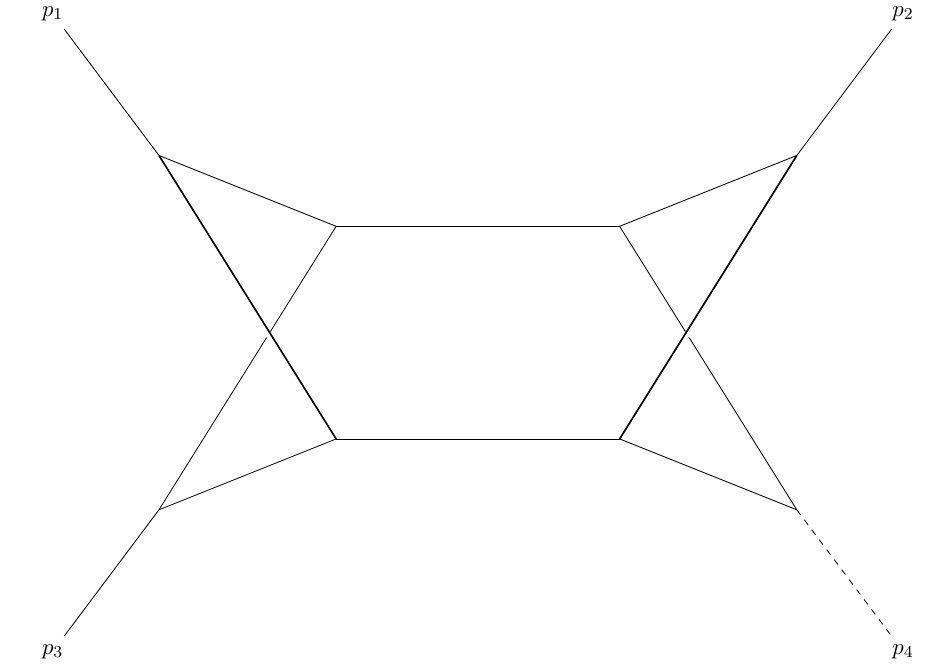}
    \caption{NPL2\_8121}
\end{subfigure}\hfill
\begin{subfigure}[b]{0.5\textwidth}
    \includegraphics[width=\textwidth]{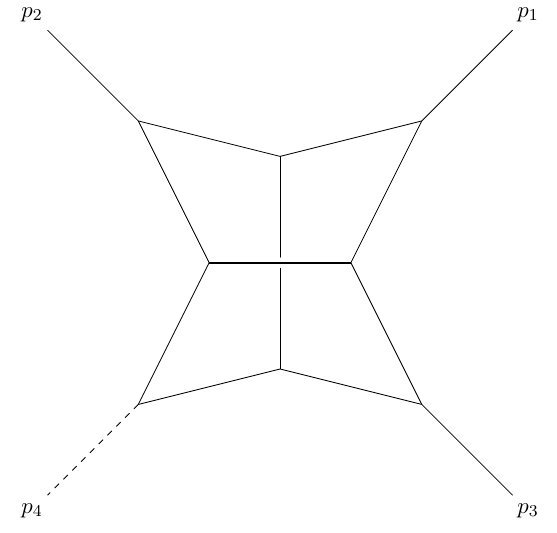}
    \caption{NPL2\_16297}
\end{subfigure}
\caption{Additional non-planar top sectors.}
\label{fig:nonpltop2}
\end{figure}

Having the canonical differential equations for topologies (\textbf{a}) and (\textbf{b}) of figure \ref{fig:nonpltop}, we can see that the former satisfies the same 6-letter alphabet \eqref{eq:pl_letters} as the planar master integrals, while for the latter we see that two new letters appear as second-order polynomials in $x~(=s_{12}/q^2)$,
\begin{equation}\label{eq:npl_letters}
    \{x,z,1-x,1-z,1-x-z,x+z,1-2x+x^2-z,x-x^2-z\}.
\end{equation}
We may nonetheless obtain a solution for topology (\textbf{b}) in terms of MPLs by integrating first over $z$ and then over $x$. Of course one needs to fix somehow the boundaries in order to obtain a final solution for the master integrals. In contrast with the planar case, here the master integrals have a more involved analytic structure, with branch points at $\{z=0,\,z=1-x,\,x=0\}$. Regularity constraints therefore are not enough to fix all necessary boundary constants. We followed a more universal approach, where we studied the solution of the differential equations to all physical and unphysical limits \cite{Henn:2020lye,Wasser:2022kwg}. The $\mathrm{d}\log$ form  of \eqref{eq:dlog_de} allows us to write its solution at the limit $\alpha_i = 0$ formally as
\begin{equation}\label{eq:bounds}
    \exp\{\epsilon B_i \log(\alpha_i)\}\vec{\mathrm{g}}|_{\alpha_i = 0} = C_i  \vec{\mathrm{g}}|_{\alpha_i = 0}
\end{equation}
The matrices $C_i$ contain terms of $\alpha_i^{n_i \epsilon}$, with $n_i$ being eigenvalues of $B_i$. We proceed then by imposing that unphysical singularities $\{z=-x,z=1,x=1\}$, i.e. terms involving $\alpha_i^{n_i \epsilon}$ with $n_i\neq 0$, must vanish at $\vec{\mathrm{g}}|_{\alpha_i = 0}$. This is another way of imposing the usual regularity conditions. For physical singularities $\{z=0,\,z=1-x,\,x=0\}$, we impose that terms involving $\alpha_i^{n_i \epsilon}$ with $n_i > 0$ must vanish at $\vec{\mathrm{g}}|_{\alpha_i = 0}$. This choice is justified by the fact that the canonical basis are free of UV divergencies, as well as by checking using expansion-by-regions \cite{Jantzen:2012mw} that for the specific limits the scaling in powers of $\epsilon$ never comes with a positive sign. Another way of arguing about this specific choice of sign is by observing that the two-point master integrals that appear in subsectors of topologies (\textbf{a}) and (\textbf{b}) of figure \ref{fig:nonpltop}, e.g. three-loop bubbles, and are divergent at $\alpha_i = 0$, always scale as $(\alpha_i)^{(n_k \epsilon)}$, with $n_k < 0$. This procedure allowed us to fix all necessary boundary constants and obtain analytic results for the two non-planar topologies of figure \ref{fig:nonpltop} in terms of MPLs up to transcendental weight six. We have cross checked our results against \cite{Henn:2023vbd}, finding perfect agreement.



\begin{figure}[t]
    \centering
    \includegraphics[width=5cm]{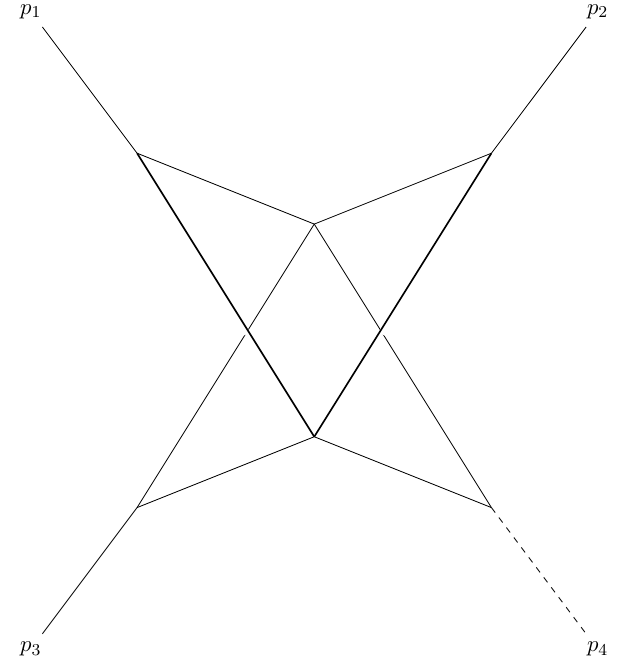}
    \caption{Subsector with square root leading singularity.}
    \label{fig:sqrt}
\end{figure}

Before concluding, we would like to comment on two other non-planar top sectors shown in figure \ref{fig:nonpltop2}. Topology (\textbf{a}) consists of 121 master integrals, with 3 of them at the top sector, while for topology (\textbf{b}) we have 371 master integrals, with 19 of them at the top sector. Regarding topology (\textbf{a}), we have observed that one of its subsectors, shown in figure \ref{fig:sqrt}, has a square root involving all kinematic invariants, $\sqrt{q^2\,s_{12}\,s_{23}\,s_{13}}$, as leading singularity. Although this square root is rationalizable, allowing for a solution of the relevant canonical differential equations in terms of MPLs, special care has to be taken when considering the analytic continuation to physical regions of phase space. For topology (\textbf{b}), constructing a canonical basis of master integrals is work in progress. The $19\times19$ system of differential equations at the top sector poses a significant challenge, however we are confident that a solution for this highly non-trivial topology will be achieved in the foreseeable future. 

\section{Conclusions}
In these proceedings we presented results on the three-loop master integrals relevant for processes such H+jet and Z+jet production at N$^3$LO. We provided a complete description of all planar master integrals and discussed the analytic computation of two non-planar topologies in terms of MPLs. Finally, we commented on the ongoing effort to compute some additional non-planar topologies. Results for the planar family and the two non-planar topologies (\texttt{NPL2c24\_15055}, \texttt{NPL2\_15055}) are provided in ancillary files attached to the \texttt{arxiv} submission of this contribution.

\acknowledgments
The research work of DC was supported by the Hellenic Foundation for Research and Innovation (HFRI) under the HFRI Ph.D. Fellowship grant (Fellowship Number: 554). NS was supported by the Excellence Cluster ORIGINS funded by the Deutsche Forschungsgemeinschaft (DFG, German Research Foundation) under Germany's Excellence Strategy - EXC-2094 - 390783311.

\bibliographystyle{JHEP}
\bibliography{skeleton}

\end{document}